\documentclass [preprint,aps,prb]{revtex4}

\begin{document}
\begin{center}
\textbf{\Large Quantum Interference in Deformed Carbon Nanotube Waveguides}
\end{center}

\vskip 2cm

\begin{center}
Wei Fa $^{1, 2}$ and Jinming Dong$^{1}$
\end{center}

\begin{center}
\textit{$^1$ National Laboratory of Solid State Microstructures and Department of Physics, Nanjing University, Nanjing, 210093, People's Republic of China}

\textit{$^2$ Department of Applied Physics, Nanjing University of Aeronautics and Astronautics, Nanjing, 210016, People's Republic of China}
\end{center}

\vskip 2cm

\begin{center}
{\bf Abstract}
\end {center}

Quantum interference (QI) in two types of deformed carbon nanotubes (CNTs), i.e., 
axially stretched and AFM tip-deformed CNTs, has been investigated by the 
\textit{$\pi $}-electron only and four-orbital tight-binding (TB) method. 
It is found that the rapid conductance oscillation (RCO) period is very sensitive 
to the applied strains, and decreases in an inverse proportion to the deformation degree, 
which could be 
used as a powerful experimental tool to detect precisely the deformation 
degree of the deformed CNTs. Also, the \textit{$\sigma -\pi $} coupling 
effect is found to be negligible under axially stretched strain, while it works on the 
transport properties of the tip-deformed CNTs. 

\vskip 2cm


\clearpage

The mechanical and electronic properties of carbon nanotubes (CNTs) have 
drawn a great deal of interests due to the potential applications in developing further quantum 
devices.$^{1 - 3}$ A perfect CNT can 
act as either a metal or a semiconductor, depending on its helicity and 
diameter.$^{4, 5}$ However, various mechanical deformations may exist in 
the actual CNTs due to their experimental surroundings, which can bring 
substantial changes of their electronic structures.$^{6 - 8}$ For examples, 
a metal-semiconductor transition occurs under uniaxial strain for all 
metallic CNTs except armchair ones, and the change rate of band gap vs 
strain increases as the chiral angle decreases. Additionally, the 
conductance of a metallic CNT could decrease by several orders of magnitude 
under the manipulation of an atomic force microscope (AFM) tip, presenting a 
possible application of the CNTs in nano-engineering, such as 
nanoscale sensors and electromechanical switches, on the basis of the 
controllable fabrications of the deformed CNTs. Though intensive studies 
have been carried out to understand the deformed CNTs, there still exist 
many experimental and theoretical uncertainties about the intimate 
relationships between their electrical and mechanical properties. Also, how 
to measure the deformation degree remains a major challenge to the existing 
experimental techniques.

It is well-known that the quantum interference (QI) between electron waves 
becomes increasingly crucial as electronic devices shrink down to nanometer 
size.$^{9 - 11}$ The rapid and slow conductance oscillations (RCO and SCO) of the metallic CNT electron resonators 
have been observed to be dependent on the tube length and 
chiral angles.$^{12,13}$ Jiang \textit{et al.} have derived an analytical expression
and successfully explained the above experimental results, showing that both the RCO and SCO are induced by the intrinsic QI.$^{14}$ However, the effects of 
deformation have not been considered in their work.
Just as the optical interference plays an important role in 
modern precise measurements, the QI of electron waves may 
be also useful in detecting precisely nano-size structures. 

Therefore, in this paper, we first present our numerical simulations on two 
types of deformed CNTs, i.e. axially stretched and AFM-deformed CNTs. We 
find that the \textit{$\sigma -\pi $} hybridization effect is negligible on the electronic 
properties of the axially stretched CNTs, but works on those 
AFM-deformed CNTs. The most interesting finding is that the RCO period decreases as the increase of deformation in 
both cases, which may provide a powerful experimental tool to measure 
precisely the deformation degree of the deformed CNTs.

The whole model system consists of a central CNT and two semi-infinite leads 
(left and right), which, for simplicity, are taken to be the same tubes as 
the central sample. We have studied two kinds of different chiral CNTs in 
detail, which, for comparison, the original length of the central sample 
($L)$ is taken to be approximately equal, containing 146 unit cells for the 
armchair tube (6, 6) and 84 cells for the zigzag one (12, 0), respectively. 
For simplicity, we label the $(n, m)$ CNT with $k$ cells as $(n, m) k$ throughout the paper. 

Atomic simulations are carried out on our samples by 
considering two types of deformation, i.e., stretched tubes under uniform 
uniaxial strain and tubes pushed by an atomically sharp AFM tip. To simulate 
the former, the tube is firstly stretched rigidly by a distance $\Delta L$ 
along its axis with one side fixed. Then, atoms at both sides of the central 
sample, contacting with the leads, are fixed and the remaining structure is 
relaxed with the universal force field (UFF).$^{15}$ As shown in Fig. 1, for the tip 
deformation, a 15-atom Li-needle normal to the (100) direction is chosen to 
model the AFM tip as in Ref. 8, which is first aimed at the center of a hexagon 
on the bottom side of the midder part and then pushed downward vertically by a displacement \textit{$\delta $} to bend the 
CNT, resulting in the total length of the deformed tube to be $L + \Delta L 
= \sqrt {L^2 + 4\delta ^2} $. The whole inflected tube is then relaxed by 
UFF, keeping the needle atoms and the end atoms of both sides fixed. Since 
the external stress is locally concentrated in the middle part adjacent 
to the Li-tip, a structural optimization based on density functional theory (DFT) 
is further applied to relax a smaller 
240-atom section in the middle. The DFT code used here is Accelrys' 
DMol3,$^{16}$ in which the electronic wave functions are expanded in a 
double-numeric polarized (DNP) basis set with a real-space cutoff of 4.0 
{\AA}. Approximation used in the Hamiltonian is the Harris functional$^{17}$ 
with a local exchange-correlation potential.$^{18}$ The inflected tube at every deformation degree
is relaxed sufficiently, and so a quasi-static process is simulated.

To compare the two types of deformation directly, we denote the deformation 
degree by the tube-length change. For example, the 5{\%} strain means 
a total tube-length increase of 5 percent, either in axially stretched or 
tip-induced deformation. The cross section holds a nearly perfect circle in 
the stretched CNT with bonds almost uniformly elongated along the tube-axis. 
In contrast, there exists a more increase of the average bond length in the 
highly deformed region of the tip-deformed tube. It is interesting that, as 
those discussed in Ref. 8, sp$^{2}$-coordinated configurations are kept in 
the action of tip, even for the largest deformation (10{\%}) 
considered here.

Then, the quantum conductance fluctuation of the deformed CNT electron resonator has been 
calculated by the tight-binding (TB) model 
including only the nearest-neighbor interactions. Firstly, we use a simple 
\textit{$\pi $}-electron approximation, in which all the hopping integrals are assumed to be 
$V_{pp\pi } = - 2.75$ eV$^{19}$ multiplied by a scaling factor $(r_0 / r_i 
)^2$, where $r_0$ and $r_i$ are bond lengths without and with 
deformation, respectively.$^{20}$ In order to simulate the boundary effect, for 
those atoms at contacts, their hopping parameters are taken to be $\alpha 
\cdot V_{pp\pi }$ with $0 < \alpha < 1$.
Consequently, electrons will be 
slightly scattered at the interfaces for $\alpha \ne 1$ and the whole system 
behaves like a Fabry-Perot electron waveguide, whose conductance $G$ is 
calculated by the Landauer formula $G = \frac{2e^2}{h}T$, where $T$ is the 
transmission coefficient. Since the transmission probability $T$ reaches to 
$0.5 \sim 0.95$, as shown in the experiment,$^{13}$ the $\alpha$ value is resonably set to be
$0.7$ in our simulations.$^{21}$ 

Four-orbital calculations are also performed to show the \textit{$\sigma -\pi$} coupling effect. 
The hopping integrals are taken to be close to those used for 
graphite,$^{22}$ i.e., $V_{ss\sigma } = - 4.43$ eV, $V_{sp\sigma } = 4.98$ eV, 
$V_{pp\sigma } = 6.38$ eV, and $V_{pp\pi } = - 2.66$ eV, which have been 
successfully used in the SWNT with small radius.$^{23}$ Among the four 
orbitals per atom, its $s$ level is located at $\varepsilon _s^0 = -7.3$ eV 
below the triply-degenerated $p$ level taken as the zero of energy. Like the 
\textit{$\pi $}-electron calculations, the same bond-length-dependence of the 
nearest-neighbor hopping parameters is taken into account. 

The results obtained by using a single \textit{$\pi$}-orbital are discussed firstly.
We calculate conductance of two perfect-CNT electron resonators, 
which will be used as benchmarks for analyzing the QI of 
the deformed tubes. Obtained results as a function of Fermi Energy are shown in Fig. 2, from which the 
general characteristics can be clearly found. Both of the CNT electron 
resonators, $(6, 6) 146$ and $(12, 0) 84$, display 
remarkable RCO with their maximum conductance approaching 
2$G_{0}$, but a SCO background is present obviously in the 
armchair tube and absent for the zigzag one. As interpreted in Ref. 14, both 
the RCO and SCO are the QI phenomena, 
which come from the linear and nonlinear terms in the energy 
dispersion relations, respectively. The zigzag CNT resonator has no SCO due to 
existence of identical energy dispersion 
relations for its two propagating modes. The calculated RCO period 
$\Delta E_F$ equals to $51.54$ meV for (6, 6)146 and 
$52.12$ meV for (12, 0)84, which agree very well with the theoretical 
prediction $\Delta E_F = hV_F / 2L \approx 18.3926 / L$ ($L$ is the 
tube-length in unit of {\AA}). In addition, a small band gap exists in the 
zigzag tube (12, 0) due to curvature, induced by the bond-dependent hopping 
integrals.

Then, we consider effects of the two types of deformation on the conductance 
oscillations and the corresponding results are illustrated in Figs. 3 and 
4. For clarity, we just show the cases of deformed tubes with length increases 
of 5{\%} and 10{\%} in Fig. 3. 
Application of uniaxial strain opens a band gap only for the zigzag tube, 
which is coincident with the conclusion of Ref. 6. Of special interest is 
that both the RCO and SCO are preserved well for the 
stretched CNTs up to axial strain of 10{\%}. Moreover, it is obvious that 
the RCO period decreases with increase of deformation, which is 
illustrated by its frequency shift to the higher one, shown in the 
inset of Fig. 3. For example, the RCO periods of the (6, 6)146 
under strain of $5{\%}$ and $10{\%}$ equal to $48.10$ meV and $44.50$ meV, 
respectively. For the axially stretched zigzag tube, though the band gap is 
opened, the successive decrease of the RCO period is still 
distinct, since the overall stretching causes the change of the phase 
difference. It can be seen obviously from Fig. 3 that the RCO becomes faster 
near the band edges of the stretched (12, 0) tube, in which a 
higher frequency with smaller amplitude emerges in the frequency analysis 
except the intrinsic RCO frequency. Additionally, obtained 
results of the two achiral tubes show that the change rate of the RCO period 
with respect to strain ($d(\Delta E_F) / d(strain))$ 
may be also chirality-dependent (see Fig. 4).

As for the AFM tip-deformed tubes, the situation becomes more complicated. 
Although the armchair tube (6, 6) remains metallic under all conditions 
considered, there exists a large drop in its conductance around $E_{F}$ when 
the total-length increase exceeds 7{\%} due to existence of localized states 
induced by the concentrated local strain in the middle part. However, the 
RCO and SCO of the armchair tubes are still obvious in the 
tip deformation and the decrease of the RCO period is also 
observed, as illustrated in Fig. 3 (a). For the tip-deformed zigzag tube 
(12, 0), the main characteristics of QI are similar to 
those in the stretched strain. Interestingly, a new fluctuation with a half 
of the RCO frequency appears more obviously with increase of 
tip-deformed degree, which may be induced by the heavy local structural 
deformation near tip and ascribed to nonlinear effects. By plotting $\Delta 
E_F$ versus tip-deformed degree in Fig. 4, we found that the slope of 
$d(\Delta E_F) / d(strain)$ is slightly larger than that of the axially 
stretched deformation. This is not surprising because a larger tensile strain 
exists in the middle of the tip-deformed tube, while the increase of 
bond-length is almost uniform in the axially stretched CNT. 

It is clearly seen from Fig. 4 that both of our numerical results on the 
axially stretched and tip-deformed CNTs are almost quantitatively consistent with 
those from the analytical formula in Ref. 14. However, there is still a minor difference 
between them, especially for the larger strains (e.g. $>3\%$), caused by a 
non-uniform change of the bond lengths in the present 
finite-length CNTs. That is to say, both the total 
tube-length and the bond-length changes can affect the RCO. 

Furthermore, four-orbital TB calculations are performed to take into account any 
effect associated with \textit{$\sigma -\pi $} hybridization. For axially stretched deformation, 
the main results about the RCO period agree with the \textit{$\pi $}-orbital 
results presented in Figs. 3 and 4: the RCO period change is 
inversely proportional to the deformation degree. The difference is that the 
slope of $d(\Delta E_F) / d(strain)$ is slightly steeper than that of the 
\textit{$\pi $}-orbital results, which means the effect of the stretched bond-length on 
$\Delta E_F$ is more distinct due to \textit{$\sigma -\pi $} hybridization. 

However, the \textit{$\sigma -\pi $} coupling has actual effects on the transport properties 
of the tip-deformed CNTs. Taking the tip-deformed (6, 6)146 as an example, 
as illustrated in Fig. 5, the SCO can survive in 
the case of smaller tip-deformation ($<6{\%}$), but will be smeared out as the 
strain becomes larger. Additionally, it is also seen from Fig. 5 that the 
localized states have more important influence on the system conductance 
than the corresponding \textit{$\pi $}-orbital results. For example, the concentrated local 
strain induces a conductance dip above $E_{F}$ for 5{\%} strain, in which a 
localized state at 0.68 eV appears. However, though the SCO 
gradually disappears with increase of the tip-deformation degree, the RCOs still 
keep in every case. More importantly, the $\Delta E_F$ 
decreases gradually as the tip pushes, and the main frequency shifts to a 
higher one. As a result of the stretched tube-length, the RCO 
period does decrease with increase of deformation degree in both cases, 
i.e. the axial strain and tip-induced deformation.

In addition, we have also calculated other finite-length tubes with 
different chiralities, such as (8, 2)55 and (6, 3)32, and found that chiral 
tubes satisfying $n-m$$=3*integer$ have similar results as those of the (12, 0)84. 
Thus we can conclude that our conclusions obtained above are suitable for 
the QI of the metallic deformed-CNT electron resonators. 

In conclusion, we have shown that the two types of mechanical deformations 
can significantly modify the QI of CNTs and the RCO periods 
are very sensitive to the applied strains. 
It is found that although there are some differences in the QI of the two 
types of deformed CNTs, the RCO period change depends 
mainly on the tube-length and is inversely proportional to the strain. Since the Fermi energy change 
$\Delta E_F$ is driven by the gate voltage $V_g$ as $\Delta E_F = C_{eff} \cdot \Delta V_g$ (typically, 
the gate efficiency factor $C_{eff}$ = 
0.01 $\sim$ 0.05), a measurement of the RCO can offer a 
possible powerful tool to measure precisely the deformation 
degree of the deformed CNT. Application of such techniques may therefore open a door 
in future to use a metallic CNT as a tiny electro-mechanical switch under 
the effects of various deformations.

This work was supported by the Natural Science Foundation of China under 
Grants Nos. A040108 and 90103038.

\clearpage

\begin{center}
{\bf Figure Captions}
\end{center}

\vskip 0.5cm

\noindent FIG. 1. The CNT inflected by an AFM 
tip. (a) Schematic view; (b) Relaxed structure of the tip-deformed tube (12, 0) with a length increase of 
7 percent, shown only 240 atoms in the middle (further 
relaxed by DFT). The Li tip (white balls) and ending atoms (black balls) are fixed at their 
respective classical positions in the DFT optimization. 

\vskip 0.5cm

\noindent FIG. 2. TB \textit{$\pi $}-electron results on conductance $G$ (in unit of $G_{0})$ vs 
Fermi energy \textit{E}$_{F}$ with \textit{$\alpha $}=0.7 for perfect CNTs: (a) (6, 6)146 
and (b) (12, 0)84.

\vskip 0.5cm

\noindent FIG. 3. TB \textit{$\pi $}-electron results on conductance $G$ vs 
Fermi energy \textit{E}$_{F}$ with 
\textit{$\alpha $}=0.7 for the deformed tube: (a) (6, 6)146 and (b) (12, 0)84. For clarity, 
only 5{\%} and 10{\%} strain are shown for the axially stretched and 
tip-induced deformation, denoted by ``S'' and ``Tip'', respectively. The 
inset in each panel is the corresponding frequency analysis for the 
conductance oscillation, in which the dotted line represents the oscillation 
frequency of the perfect tube. 

\vskip 0.5cm

\noindent FIG. 4. (Color online) RCO period change of the deformed tubes, obtained from 
a TB\textit{ $\pi $}-electron calculation. The dashed line corresponds to the analytical 
result for the axially stretched tube (6, 6)146. 

\vskip 0.5cm

\noindent FIG. 5. Four-orbital TB results for the perfect (6, 6)146 and its 
tip-deformed cases under strain of 5{\%} and 10{\%}. The left panel gives 
the conductance $G$ vs Fermi energy \textit{E}$_{F}$ with \textit{$\alpha $}=0.7, and the right panel shows 
the corresponding frequency analyses. For comparison, the \textit{$\pi $}-electron 
frequency analyses are also shown by the dashed lines.

\end{document}